\documentclass[twocolumn,aps,amssymb,prb]{revtex4}
\usepackage{graphicx}
\usepackage{epsfig,amsmath}

\begin{document}

\title{Quantum squeezing by a parametric resonance in a SQUID}

\author{T.~Ojanen$^1$}
\email[Correspondence to ]{teemuo@boojum.hut.fi}
\author{J. Salo$^2$}
 \affiliation{ $^1$Low Temperature Laboratory, Helsinki University of
Technology, P.~O.~Box 2200, FIN-02015 HUT, Finland }
\affiliation{$^2$Laboratory of Physics, Helsinki University of
Technology, P.~O.~Box 4100, FIN-02015 HUT, Finland }
\date{\today}

\begin{abstract}
We study rotating squeezed quantum states created by a parametric
resonance in an open harmonic system. As a specific realization of
the phenomenon we study a mesoscopic SQUID loop where the state
preparation procedure is simple in principle and feasible with
currently available experimental methods. By solving dynamics and
calculating spectral properties we show that quantum fluctuations of
SQUID observables can be reduced below their groundstate value. The
measurement is introduced by coupling the SQUID to a transmission
line carrying the radiation to a secondary measurement device.
Besides the theoretical interest, our studies are motivated by an
opportunity for a practical quantum noise engineering.
\end{abstract}
\pacs{PACS numbers: } \bigskip

\maketitle

\section{Introduction}
At the heart of the quantum theory lies the fundamental principle of
describing physically observable quantities as Hermitian operators
acting on Hilbert space of quantum states. Generally these operators
do not commute, a trait giving rise to the fundamental uncertainty
principle first formulated by Heisenberg. The uncertainty principle
characterizes the statistical spread of the distributions
corresponding to two different observables. If the operators
representing the observables do not commute, the statistical
variations of their observed distributions, frequently called
uncertainties, cannot generally be arbitrarily small in a given
state. However, the statistical variation of a single observable is
not limited in any way by the uncertainty principle. For two given
non-commuting observables, a state in which the lower limit of the
uncertainty principle is reached is called a minimum uncertainty
state. Decreasing the uncertainty with respect to one observable
results in the increase of the uncertainty of the other. The
transfer of uncertainty from one observable to another with respect
to the minimum uncertainty values is referred to as squeezing. The
squeezing of quantum fluctuations was first studied and
experimentally verified in quantum optics, where the components of
quantized electric field served as the squeezed
observables.\cite{slusher,loudon} Since then the phenomenon has been
observed in superconducting circuits,\cite{yurke1,yurke} and more
recently, there has been a promising efforts to realize the
squeezing in nanomechanical structures.\cite{ruskov}

Due to the technological advances, the development in mesoscopic
physics has been rapid in last decades, and many interesting quantum
phenomena have been experimentally verified for the first time. Also
a number of new and important measuring devices, necessary to
observe elementary quantum phenomena, have been invented. In this
paper we study the phenomena of the quantum squeezing in a
mesoscopic Superconducting QUantum Interference Device (SQUID), with
emphasis on the creation and consequences of the squeezing. However,
the squeezing is an example of a manipulation of quantum
fluctuations which could have direct applications in a practical
quantum measurement.

 Experimentally the squeezing of the
quantum fluctuations has been demonstrated in  microwave frequencies
by constructing a Josephson parametric amplifier,\cite{yurke} with a
40\% reduction of the vacuum noise. Theoretically the squeezing has
been
 studied in a similar SQUID ring we
consider here.\cite{everitt} The squeezing mechanism we study is
different from one studied in Ref. \onlinecite{everitt} which was
based on a rapid decrease of an external magnetic flux to switch on
the Josephson coupling. We consider a parametric instability in a
harmonic regime,\cite{brown} a well-known procedure to create the
squeezing and applicable in various different
systems.\cite{averbukh}  It can be realized with an elementary flux
control in an
 rf SQUID. In the limit of a negligible dissipation the
magnitude of the squeezing is exponential in short times and, it
rotates between the charge and the magnetic flux of the SQUID. A
strong dissipation compensates the resonant driving and leads to a
rotating quasistationary state where uncertainties periodically go
below their groundstate values. We calculate the noise spectrum of
periodic squeezed states and discuss the measurement problem by
analyzing a setup where the SQUID is coupled to a transmission line.

\section{Squeezing by parametric resonance}

\begin{figure}[h]
\centering
\begin{picture}(100,160)
\put(-40,0){\includegraphics[width=0.8\columnwidth]{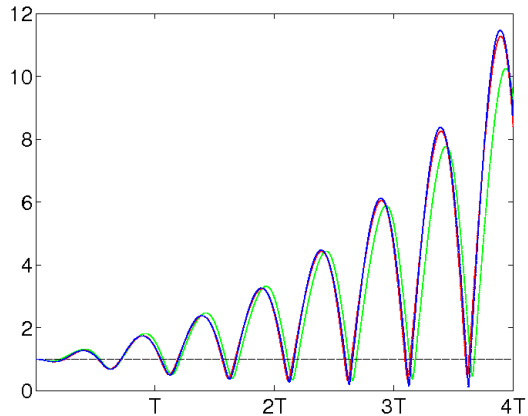}}
\end{picture}
\caption{(Color online) Uncertainty of the magnetic flux
$\sqrt{2}\Delta\phi'$ in time. The parametric driving with a
strength $B=0.1\omega_0^2$ makes the uncertainty to oscillate with
the frequency $2\omega_0$ below and above the groundstate value
marked by the black dashed line. The blue line corresponds to the
analytical solution, the red
 line to the exact numerical solution and the green
line to the exact solution with the AC flux modulation (\ref{ehto}).
 } \label{squeeze3}
\end{figure}

The parametric resonance is a well-known physical phenomena in a
classical harmonic oscillator (HA).\cite{landau} A periodic
time-dependent perturbation can lead to large effects, which are
most significant when the period of the perturbation is twice the
resonance frequency. Driven in resonance, the energy of the system
grows exponentially in time. This is true for a quantum HA
also.\cite{averbukh} In addition, starting the resonant driving from
the groundstate of HA it results in a rapid squeezing of the
uncertainties of the conjugate observables of HA.
 Consider the
Hamiltonian
\begin{equation}\label{parres}
H=\frac{p^2}{2m}+\frac{m\omega^2(t)}{2}x^2,
\end{equation}
where $p$ and $x$ are the canonical variables $[x,p]=i\hbar$, $m$ is
the mass constant and the time-dependent dependent perturbation is
included in $\omega^2(t)=\omega_0^2+A\,\mathrm{cos}\,2\omega_0t$.
Furthermore, let us assume that the time-dependent part can be
treated as a perturbation, $|A|\ll\omega_0^2$. The groundstate of
the free Hamiltonian ($A=0$) in the position representation is
$\psi_0(x)=\pi^{-1/4}\alpha^{1/4}\mathrm{exp}(-\alpha x^2/2)$, where
$\alpha=1/l_0^2=m\omega_0/\hbar$ is an inverse square of the
characteristic zero point length scale. An approximate solution to
the Schr\"odinger equation corresponding to the Hamiltonian of Eq.
(\ref{parres}) and the initial condition $\psi(x,t=0)=\psi_0(x)$ can
be obtained analytically in the  Gaussian form
\begin{equation}\label{ratk}
\psi(x,t)=\left(\frac{\mathrm{Re}
\,\alpha(t)}{\pi}\right)^{1/4}\mathrm{exp}\left(-\alpha(t)\,x^2/2\right),
\end{equation}
where
\begin{align}\label{koalpha}
\alpha(t)=\frac{i}{l_0^2}\frac{(1-ie^{2\xi})\mathrm{sin}(\omega_0t)
+(1+ie^{2\xi})\mathrm{cos}(\omega_0t) }
{-(1+ie^{2\xi})\mathrm{sin}(\omega_0t)
+(1-ie^{2\xi})\mathrm{cos}(\omega_0t) }
\end{align}
with $\xi=At/4\omega_0$. Physically Eq. (\ref{ratk}) describes a
wave packet centered at $x=0$, but whose width oscillates in time
with an exponentially growing amplitude. When the statistical spread
in the real space is the widest, the corresponding spread in the
momentum space is the smallest. Defining the dimensionless
quantities $x'=x/l_0$ and $p'=p\,l_0/\hbar$ we discover that in the
state (\ref{ratk}) we have $\Delta x'\,\Delta p'=\langle
x'^2\rangle^{1/2}\langle
p'^2\rangle^{1/2}=|\langle[x',p']\rangle|/2=1/2$ and the state
remains minimum uncertainty state at all times. The individual
deviations may be written as
\begin{equation}\label{uncer}
\Delta x'=\frac{\beta}{\sqrt{2}}\qquad \Delta
p'=\frac{\beta^{-1}}{\sqrt{2}},
\end{equation}
where $\beta=\,(\mathrm{Re}\,\alpha(t))^{-1/2}/l_0$. From Eq.
(\ref{koalpha}) we obtain
\begin{align}\label{alpha}
\mathrm{Re}\,\alpha(t)=\frac{1}{l_0^2}\left(
\mathrm{sinh}^2\xi+\right.&\left.
\mathrm{cosh}^2\xi\right.\nonumber\\
&\left.-2\,\mathrm{sin}(2\omega_0t)
\mathrm{sinh}\xi\mathrm{cosh}\xi\,\right)^{-1}
\end{align}
The maximum uncertainty oscillates with a frequency $2\omega_0$
between the position and momentum. At the times corresponding to
$\mathrm{sin}(2\omega_0t)=\pm 1$ the squeezing is given by
$\beta=e^{\mp\xi}$. The solution (\ref{ratk}) is very accurate as
long as the condition $A\ll\omega_0^2$ holds. If $A\lesssim
0.1\omega_0^2$ the analytical solution agrees excellently with the
exact solution up to the squeezing factor 5, see Fig.
\ref{squeeze3}. After that the lower value does not decrease further
as in the analytical solution even though the general agreement is
accurate.

\begin{figure}[h] \centering
\begin{picture}(100,140)
\put(-30,0){\includegraphics[width=0.6\columnwidth]{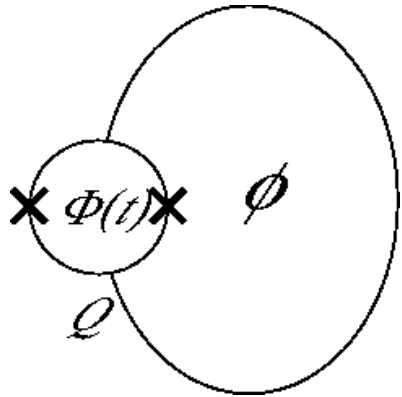}}
\end{picture}
\caption{Schematic representtion of the SQUID loop. The crosses
represent Josephson junctions and the physical quantities $\phi$ and
$Q$ correspond to the magnetic flux through the main loop and the
charge at the junctions. The flux $\Phi(t)$ through the smaller loop
is controlled by an external magnetic field.} \label{SQUID}
\end{figure}

The above example illustrates the mechanism behind the squeezing
scheme we will now apply to a flux-controlled rf SQUID loop
described in Fig. \ref{SQUID}. The Hamiltonian for the system can be
written as
\begin{equation}\label{Ham}
H=\frac{Q^2}{2C}+\frac{\phi^2}{2L_S}+E_J\mathrm{cos}(\Phi(t)\,2e/\hbar)\mathrm{cos}(\phi\,
2e/\hbar),
\end{equation}
where $C$ is the capacitance of the junctions, $L_S$ is the
self-inductance of the loop and $\Phi(t)$ is the flux bias
 externally applied through the smaller loop.\cite{makhlin} The
quantities $Q$ and $\phi$ are the canonical variables corresponding
to the charge in the capacitor and the magnetic flux  in the loop
and satisfy the commutation relation $[\phi,Q]=i\hbar$. Thus the
magnetic flux $\phi$ plays the role of the position coordinate and
the charge $Q$ corresponds to the momentum in the standard
one-particle quantum mechanics.The flux value corresponding to the
oscillator length $l_0$ of the harmonic part of (\ref{Ham}) is
$\phi_0=\sqrt{\hbar Z_0}$, where $Z_0=\sqrt{L_S/C}$.

We treat the Josephson term as a perturbation and therefore consider
 the parameter range $L_S\ll L_J$, where we have defined the
Josephson inductance as $L_J=\hbar^2/4e^2E_J$.  If the quadratic
potential is strong, the flux particle is restricted to the linear
regime of the SQUID and the term $\mathrm{cos}(\phi\, 2e/\hbar)$ may
be expanded to the second order. Supposing that the condition for
the validity of the approximation $\phi_0<\Phi_0=h/2e$ is satisfied,
the Hamiltonian (\ref{Ham}) takes the form
\begin{align}\label{Hamsquid}
  H\approx\frac{Q^2}{2C}+\frac{\phi^2}{2L_S}-&\frac{\phi^2}{2L_J}\mathrm{cos}(\Phi(t)\,2e/\hbar)=\nonumber\\
  &=\frac{Q^2}{2C}+\frac{C\omega^2(t)}{2}\phi^2,
\end{align}
where
\begin{equation}\label{taajuus}
\omega^2(t)=\omega_0^2\left(1-\frac{L_S}{L_J}\mathrm{cos}(\Phi(t)\,2e/\hbar)\right)
\end{equation}
with $\omega_0^2=(L_SC)^{-1}$. In the cosine expansion the small
term constant in $\phi$ has been dropped from Eq. (\ref{Hamsquid}).
Now we have established a connection between Eqs. (\ref{parres}) and
(\ref{Hamsquid}) provided that the external magnetic flux is
modulated according to $\Phi(t)=\hbar\omega_0t/e$. This could be
achieved by increasing the magnetic field linearly inside the
control loop. The temporal evolution of the SQUID starting from the
groundstate is readily solved by identifying
$A\to\omega_0^2(L_S/L_J) $ and $l_0\to\phi_0$
 and applying Eqs. (\ref{ratk}) and (\ref{alpha}). The
dimensionless operators $\phi'=\phi/\phi_0$ and $Q'=Q\phi_0/\hbar$
satisfy the minimum uncertainty condition $\Delta\phi'\Delta Q'=1/2$
and the individual deviations behave as those in Eq. (\ref{uncer}).

The linearly increasing magnetic flux is inconvenient practically
since the magnetic field should be restricted to the control loop
and it is difficult to isolate at high field values. In addition,
the AC fields are obtained and manipulated more easily. From the
approximate equivalence
$\mathrm{cos}\,(2\,x)\approx\mathrm{cos}\left(3\,\mathrm{sin}(x)\right)$
can be inferred that if the magnetic flux is modulated with an AC
field according to
\begin{equation}\label{ehto}
\Phi(t)=(3/2\pi)\Phi_0\,\mathrm{sin}\,(\omega_0t),
\end{equation}
then
$\mathrm{cos}(\Phi(t)\,2e/\hbar)\approx\mathrm{cos}(2\omega_0t)$.
Hence the magnetic flux should be modulated at the resonance
frequency such a way that its amplitude is roughly a half of the
flux quantum.

For later purposes it is convenient to introduce the
second-quantized creation and annihilation operators in the photon
number Fock space. The SQUID observables may be written as
\begin{align}\label{can}
\phi=\sqrt{\frac{\hbar Z_0}{2}}(a+a^{\dagger}),\qquad
Q=i\sqrt{\frac{\hbar}{2Z_0}}(a^{\dagger}-a).
\end{align}
Operators $a,\,a^{\dagger}$ satisfy the usual bosonic commutation
relations. In this language the Hamiltonian Eq. (\ref{Hamsquid}) is
transformed to
\begin{align}\label{kakkosh}
H=\hbar\omega_0(a^{\dagger}a+\frac{1}{2})+B\,
\mathrm{cos}\,(2\,\omega_0t)(a+a^{\dagger})^2,
\end{align}
where $B=\hbar\omega_0L/4L_J$. In the case where expectation values
of $\phi$ and $Q$ vanish, as above, the root mean square deviations
may be written as
\begin{align}\label{var}
&\Delta\phi'=\frac{1}{\sqrt{2}}\left(\langle a^2\rangle+ \langle
a^{\dagger 2}\rangle+\langle aa^{\dagger}\rangle+\langle
a^{\dagger}a\rangle\right)^{1/2}\nonumber\\
&\Delta Q'=\frac{1}{\sqrt{2}}\left(-\langle a^2\rangle- \langle
a^{\dagger 2}\rangle+\langle aa^{\dagger}\rangle+\langle
a^{\dagger}a\rangle\right)^{1/2}.
\end{align}
\section{Coupling the SQUID to a transmission line}

In this section we introduce a measuring scheme in which the SQUID
is coupled to a transmission line (TL) which serves as a wave guide
carrying the radiation away from the system. It also plays a role of
a generic measurement device which causes a Markovian back-action to
the measured system. In addition, the TL provides a practical
theoretical model for studying the environmental effects to the
squeezing.

The study of a non-classical radiation requires a quantum-mechanical
treatment of the SQUID and the TL. Quantum mechanics of a TL is
essentially similar with a one-dimensional quantized electromagnetic
field and it has been discussed in Ref. \onlinecite{schoelkopf}.
\begin{figure}[h]
\centering
\begin{picture}(100,100)
\put(-60,0){\includegraphics[width=0.9\columnwidth]{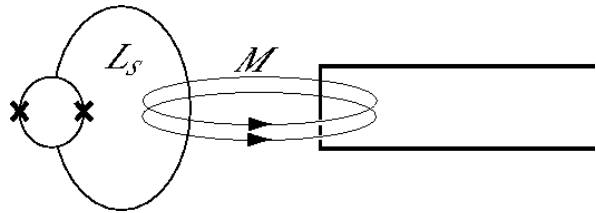}}
\end{picture}
\caption{Transmission line inductively coupled to the SQUID.}
\label{linja}
\end{figure}
Consider the system in Fig. \ref{linja} where the SQUID is
magnetically coupled to a TL. The magnetic flux of the nearby SQUID
penetrates into the TL. The inductance and capacitance per unit
length in the TL are $l$ and $c$ and the wave velocity is given by
$v=1/\sqrt{lc}$.

The Lagrangian for the line can be written as
\begin{equation}\label{Lagrange}
\mathcal{L}=\int_0^L dx\left(\frac{l}{2}j^2-\frac{c}{2}q^2\right),
\end{equation}
where $j(x,t)$ and $q(x,t)$ are the local current and charge
densities. Following Ref. \onlinecite{schoelkopf} we introduce a new
variable
\begin{equation}\label{varaus}
\theta(x,t)\equiv\int_0^{L}dx'q(x',t),
\end{equation}
which, together with the continuity equation, allows one to write
the Lagrangian
\begin{equation}\label{Lagrange2}
\mathcal{L}=\int_0^L
dx\left(\frac{l}{2}\dot{\theta}^2-\frac{c}{2}(\partial_x\theta)^2\right).
\end{equation}
 From the
total charge neutrality that it follows that $\theta(0,t)=
\theta(L,t)=0$, so $\theta$ can be expanded by the eigenmodes
\begin{equation}\label{expansion}
\theta(x,t)=\sqrt{\frac{2}{L}}\sum_k A_k(t)\,\mathrm{sin}(\beta_kx),
\end{equation}
where $b_k=k\pi/L$ and the Lagrangian takes the form
\begin{equation}\label{Lagrange3}
\mathcal{L}=\sum_k\left(\frac{l}{2}\dot{A}_k^2-\frac{\beta_k^2}{2c}A_k^2\right).
\end{equation}
The quantization of the quadratic Lagrangian (\ref{Lagrange3}) is
well known. The Hamiltonian can be written as
\begin{equation}\label{hamilton1}
H^{\mathrm{TL}}=\sum_k\left(\frac{l}{2}\Pi_k^2+\frac{\beta_k^2}{2c}A_k^2\right),
\end{equation}
where $[A_k,\Pi_k]=i\hbar$. This can be further written with the
help of bosonic creation and annihilation operators $c_k$ and
$c_k^{\dagger}$ as
\begin{align}\label{can}
&A_k=\sqrt{\frac{\hbar\omega_kc}{2}}\frac{1}{\beta_k}(c_k(t)+c_k^{\dagger}(t))\nonumber\\
&\Pi_k=-i\sqrt{\frac{\hbar\omega_kl}{2}}(c_k(t)-c_k^{\dagger}(t)),
\end{align}
and the Hamiltonian (\ref{hamilton1}) becomes
\begin{equation}\label{hamilton2}
H^{\mathrm{TL}}=\sum_m\hbar\omega_k (c_m^{\dagger}c_m+1/2).
\end{equation}
The eigenfrequencies are defined as $\omega_k=\beta_kv$. The voltage
and current operators in the TL can be derived from $\theta$ as
$V(x,t)=(1/c)\partial_x\theta(x,t)$ and $I(x,t)=\dot{\theta}(x,t)$.
The magnetic interaction between the SQUID and the TL couples the
operators $I_S=\phi/L_S$ and $I(x,t)$, where $I_S$ and $L_S$ are the
current and the inductance of the SQUID loop. Supposing that the
SQUID couples to each mode approximately as strongly, the
interaction may be written as
\begin{equation}\label{inter}
H_{\mathrm{int}}=M\frac{\phi}{L_S}\sum_ki\sqrt{\frac{\hbar\omega_k}{Ll}}
(-c_k+c_k^{\dagger}),
\end{equation}
where M is the mutual inductance between the SQUID and the field
mode.

 Now we could proceed to two directions. On one hand we could
study the SQUID Hamiltonian by trying to eliminate the transmission
line variables (and other possible environment modes). On the other
hand we are interested in the TL output radiation to measure the
SQUID. The first problem leads to a master equation description of
the SQUID and it will be studied in the next section. The latter
problem will be pursued here.

From the Heisenberg equation of motion for the operators $c_k$ we
obtain
\begin{equation}\label{hei1}
\partial_t c_k=-i\omega_kc_k+\frac{M}{L_S}\sqrt{\frac{\omega_k}{\hbar Ll}}
\phi
\end{equation}
which can be solved as
\begin{align}\label{hei2}
 c_k(t)=&c_k(0)e^{-i\omega_kt}+\nonumber\\
 &\frac{M}{L_S}\sqrt{\frac{\omega_k}{\hbar Ll}}
e^{-i\omega_kt}\int_0^t\phi(t')e^{i\omega_kt'}dt',
\end{align}
where $\phi(t)$ is the flux operator of the SQUID. From this
expression one can see that the TL problem can be solved as soon as
the the solution to the associated SQUID problem is known. The flux
$\phi(t)$ is the only SQUID variable entering to the expression
(\ref{hei2}) and can be solved independently provided that the TL
variables can be successfully eliminated from the SQUID dynamics.
Substituting the expression ($\ref{hei2}$) into the formula for the
voltage of the TL
\begin{equation}\label{voltage}
V(x,t)=\sum_k\sqrt{\frac{\hbar\omega_k}{Lc}}\mathrm{cos}\frac{k\pi
x}{L}[c_k(t)+c^{\dagger}_k(t)],
\end{equation}
we get
\begin{equation}\label{voltage2}
V(x,t)=V_0(x,t)+\frac{M\omega_0}{\pi L_S}\phi(t-x/v),
\end{equation}
where $V_0(x,t)$ is the voltage operator of the free TL and the
second term is proportional to the retarded SQUID field. Deriving
Eq. (\ref{voltage2}) we have assumed that the level separation in
the TL is much smaller than $\omega_0$, so the TL will act as a
waveguide and not as a resonator. The current of the TL has a
similar expression.

\section{The role of the dissipation and decoherence}
\subsection{Dissipative state evolution of the SQUID}
So far we have treated the SQUID as an ideal system without any
dissipation and decoherence effects. Real mesoscopic systems are
much larger than the systems usually studied in atomic and molecular
physics and generally need to be considered as open quantum systems.
Also an important source of decoherence is the quantum measurement
process in which the system is necessarily coupled to the external
world. A state of an open quantum system can no longer be
represented by simply a state vector making it necessary to
introduce the density operator. The dissipation, energy absorption
and suppression of quantum coherence effects can be naturally
discussed within the density operator approach.

 Assuming that the level separation
 in the SQUID loop is significantly higher than the
temperature of the environment $\hbar\omega_0>k_{\mathrm{B}}T$, the
dominating environmental effect is the spontaneous emission. The
dynamics of the density operator after the Born-Markov elimination
of the electromagnetic field modes can be described by a Lindblad
master equation \cite{carmichael} of the form
\begin{align}\label{lind}
\partial_t\rho=-\frac{i}{\hbar}[H,\rho]+\kappa(2a\rho a^{\dagger}-a^{\dagger}a\rho
-\rho a^{\dagger}a),
\end{align}
where the first term is responsible for the Hamiltonian evolution by
the operator (\ref{kakkosh}) and the second term describes the
spontaneous emission into the TL, which is assumed not to bring any
signal back to the SQUID. The coefficient $\kappa$ is related to the
quality factor of the circuit by $\kappa=\omega_0/Q$. In the case of
a free SQUID the $Q$-factor can be as high as $10^3-10^4$. The
coupling to the TL should decrease $Q$ significantly to guarantee an
efficient measurement of the SQUID. The value of kappa can be
estimated then as $\kappa\approx (M/L_S)^2Z_0/Z_{\mathrm{TL}}$,
where $Z_{\mathrm{TL}}=\sqrt{l/c}$.  The disadvantage of decreased
$Q$ is the increased decoherence which is hostile to the squeezing
process.

\subsection{Expectation values}
Starting from Eq. (\ref{lind}) it is possible to derive the
equations of motion for the expectation values of the creation and
annihilation operators,
\begin{align}\label{yksi}
\partial_t\langle a\rangle=&\left(-i\omega_0-\kappa-2iB\mathrm{cos}\,(\,2\omega_0t)
\right)\langle a\rangle\nonumber\\
&-2iB\mathrm{cos}\,(\,2\omega_0t)\langle a^{\dagger}\rangle,\nonumber\\
\partial_t\langle a^{\dagger}\rangle=&\left(+i\omega_0-\kappa+2iB\mathrm{cos}\,(\,2\omega_0t)
\right)\langle a^{\dagger}\rangle\nonumber\\
&+2iB\mathrm{cos}\,(\,2\omega_0t)\langle a\rangle,
\end{align}
as well as for those of the quadratic operators
\begin{align}\label{systeemi}
 &\partial_t\langle
a^{\dagger}a\rangle=2iB\mathrm{cos}\,(2\omega_0t)\left(\langle a^2
\rangle-\langle a^{\dagger 2}\rangle\right) -2\kappa\langle
a^{\dagger }a\rangle\nonumber\\
&\partial_t\langle a^2\rangle=-2i\omega_0\langle
a^2\rangle-iB\mathrm{cos}\,(2\omega_0t)\times\nonumber \\
&\qquad\qquad\left(4\langle a^2\rangle+4\langle
a^{\dagger}a\rangle+2\right)-2\kappa\langle a^2\rangle\nonumber\\
&\partial_t\langle a^{\dagger 2}\rangle=2i\omega_0\langle
a^{\dagger 2}\rangle+iB\mathrm{cos}\,(2\omega_0t)\times\nonumber \\
&\qquad\qquad\left(4\langle a^{\dagger 2}\rangle+4\langle
a^{\dagger}a\rangle+2\right)-2\kappa\langle a^{\dagger 2}\rangle.
\end{align}
These two present a closed set of equations. The missing combination
$\langle aa^{\dagger}\rangle$ can be deduced from the commutation
relation $[a,a^{\dagger}]=1$. By solving the above equation system
we obtain the uncertainties (\ref{var}). The nature of solutions to
the system (\ref{systeemi}) depends on the relative magnitudes of
$\kappa$ and $B$. If $\kappa>B$, the dissipation will eventually
compensate the resonant driving and the solutions are periodic and
bounded. Otherwise the expectation values grow increasingly in time.
This, however, does not imply increasing squeezing of quantities
(\ref{var}). In the presence of dissipation the lowest value of the
fluctuations has a finite lower limit.

\begin{figure}
\centering
\begin{picture}(120,120)
\put(-30,0){\includegraphics[width=0.7\columnwidth]{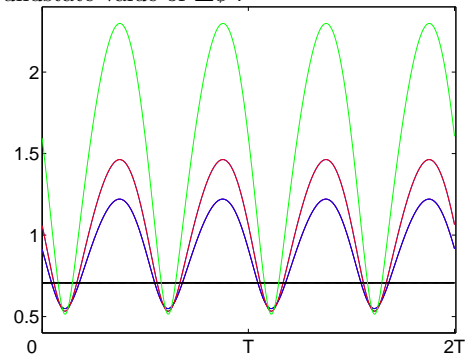}}
\end{picture}
\caption{(Color online) Uncertainty $\Delta \phi'$ of the bounded
periodic solutions $\kappa=1.5B$ (blue), $\kappa=1.3B$ (red) and
$\kappa=1.1B$ (green). The black horizontal dashed line marks the
groundstate value of $\Delta \phi'$. The minimum of the squeezing in
the periodic solutions approaches to lower bound of about 0.75 times
the groundstate value. } \label{fluksiep}
\end{figure}

By expanding the expectation values in Eq. (\ref{systeemi}) to a
Fourier series and truncating them to the fourth order, we found an
accurate analytical form for the periodic solution. In the limit
$\kappa\to B+0$, the lower limit of the periodic squeezing is about
0.75 times the vacuum value of $\Delta\phi'$ and $\Delta Q'$, see
Fig. \ref{fluksiep}.
\begin{figure}
\centering
\begin{picture}(120,120)
\put(-30,0){\includegraphics[width=0.7\columnwidth]{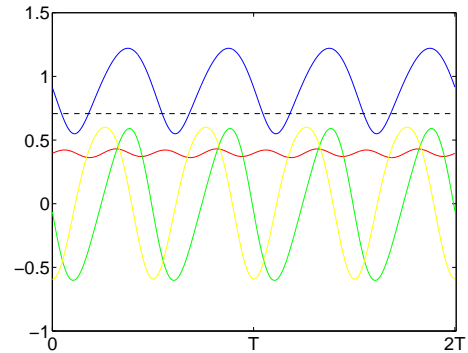}}
\end{picture}
\caption{(Color online) Two periods of the stationary solution
$\kappa=1.5B$. The blue curve is $\Delta \phi'$ and the red curve
$\langle a^{\dagger}a\rangle$. The yellow and the green curves
correspond to
 $\mathrm{Re}\,\langle a^2\rangle$ and $\mathrm{Im}\,\langle a^2\rangle$
 respectively. The black horizontal dashed line marks the groundstate value of
 $\Delta \phi'$.} \label{kaikki}
\end{figure}

\begin{figure}
\centering
\begin{picture}(120,120)
\put(-30,0){\includegraphics[width=0.7\columnwidth]{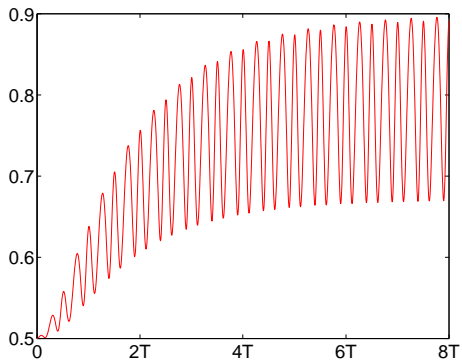}}
\end{picture}
\caption{(Color online) Evolution of the uncertainty product
$\Delta\phi'\Delta Q'$ from the groundstate to  the periodic steady
state $\kappa=1.5B$. After reaching the steady state, the
uncertainty product oscillates with a frequency $4\omega_0$ about
the constant value.} \label{unpro}
\end{figure}

The nature of squeezed states in the presence of the dissipation is
somewhat different from the pure squeezed states. The pure squeezed
states are linear combinations of the even photon number states. The
dissipation changes the spectral content so that the odd number
photon states are also occupied. The resulting state is no longer a
minimum certainty state, see Fig. \ref{unpro}.

\subsection{Two-time correlation functions}
In order to find the spectral properties of the signal radiated into
the TL, we also need to evaluate two-time correlation functions,
such as $\langle a^{\dagger}(t_2)a(t_1)\rangle$. According to the
quantum regression formula \cite{carmichael}, the function pair
$\langle A(t)a(t')\rangle$ and $\langle A(t)a^{\dagger}(t')\rangle$
obey the same differential equations than $\langle a(t')\rangle$ and
$\langle a^{\dagger}(t')\rangle$ for arbitrary operator $A(t)$.
Choosing $A(t)$ to $ a(t)$ we get
\begin{align}\label{qreg}
\partial_{t'}\langle a(t)a(t')\rangle=&\left(-i\omega_0-\kappa-2iB\mathrm{cos}\,(\,2\omega_0t')
\right)\langle a(t)a(t')\rangle\nonumber\\
&-2iB\mathrm{cos}\,(\,2\omega_0t')\langle a(t)a^{\dagger }(t')\rangle,\nonumber\\
\partial_{t'}\langle a(t)a^{\dagger }(t')\rangle=&\left(+i\omega_0-\kappa+2iB\mathrm{cos}\,(\,2\omega_0t')
\right)\langle a^{\dagger }(t)a(t')\rangle\nonumber\\
&+2iB\mathrm{cos}\,(\,2\omega_0t')\langle a(t)a(t')\rangle
\end{align}
for $t'>t$. A similar pair of equations hold for $\langle
a^{\dagger}(t)a(t')\rangle$ and $\langle
a^{\dagger}(t)a^{\dagger}(t')\rangle$. Solving Eqs. (\ref{qreg}) and
the corresponding equations to the other pair, we can deduce an
arbitrary second-order correlator in the whole time domain by using
symmetries of the correlators.

\subsection{Wigner function of the squeezed state}
\begin{figure}
\centering
\begin{picture}(100,150)
\put(-20,0){\includegraphics[width=0.5\columnwidth]{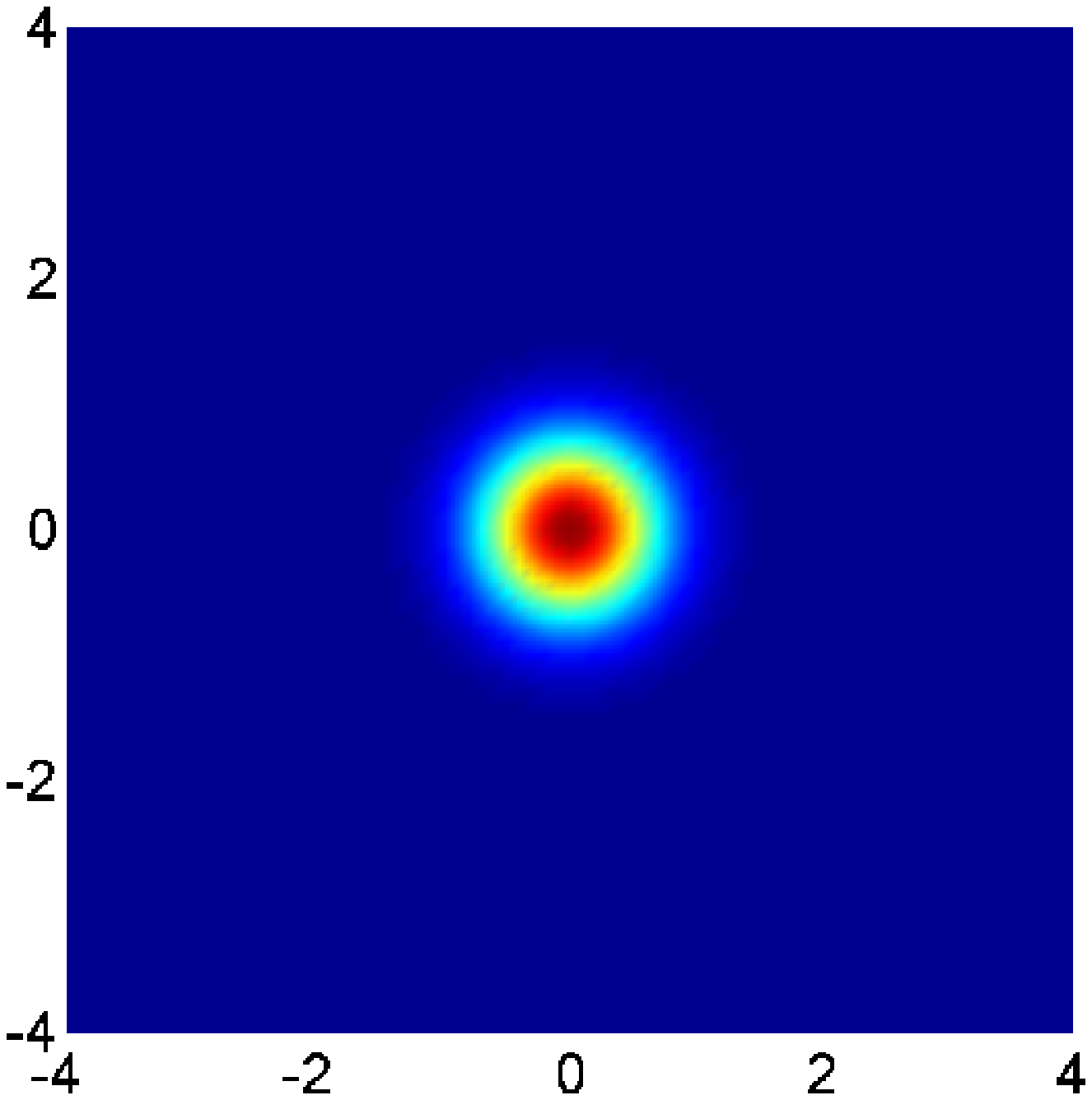}}
\end{picture}
\begin{picture}(100,100)
\put(6,0){\includegraphics[width=0.5\columnwidth]{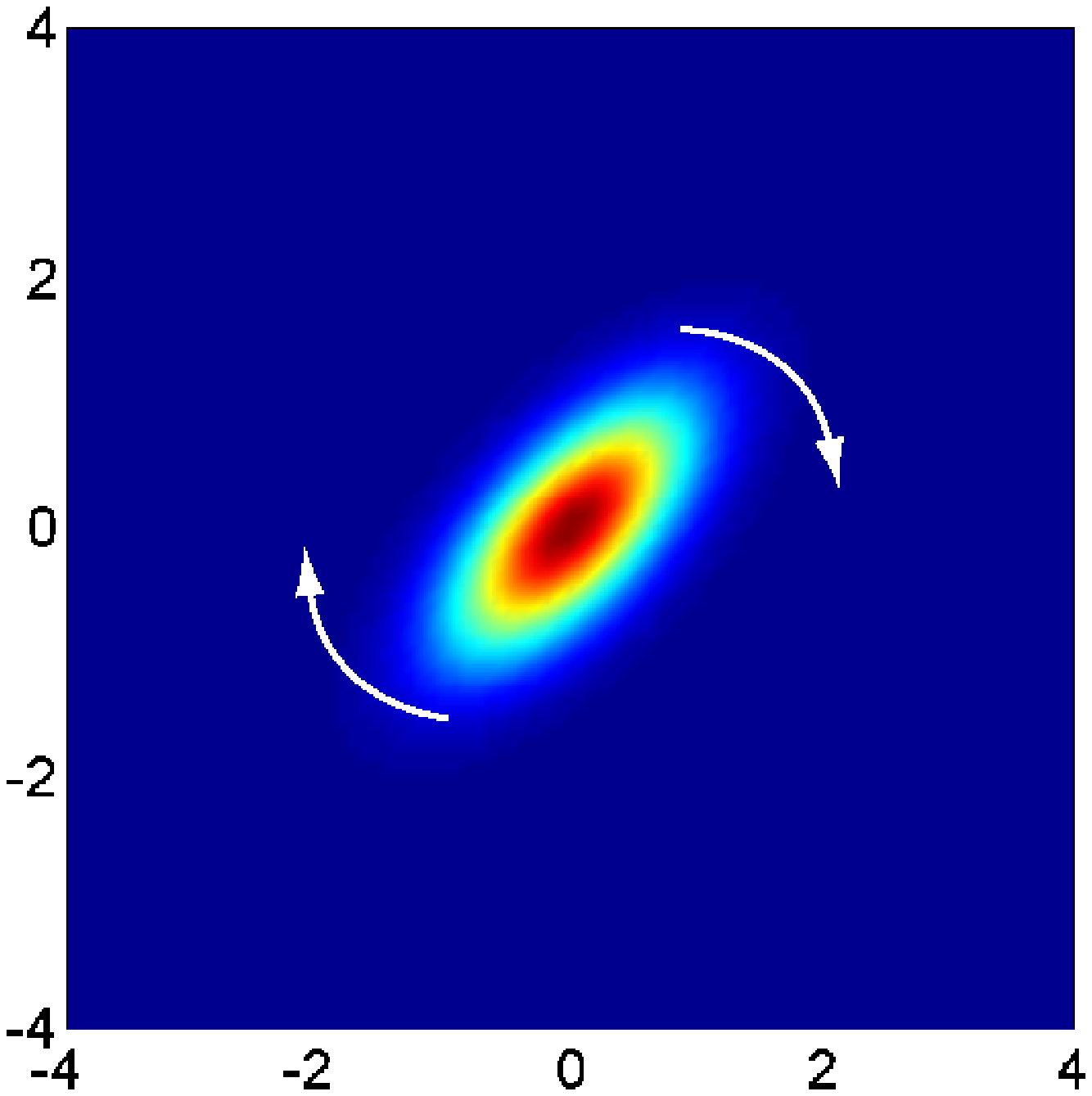}}
\end{picture}
\caption{(Color online) Wigner function of the groundstate (left)
and the periodic squeezed state $\kappa=1.5B$ (right) in the
$(\phi', Q')$-plane. The squeezed state rotates clockwise as
indicated by the arrows. The ellipse makes $2\pi$ rotation in time
$T=2\pi/\omega_0$. } \label{wig1}
\end{figure}

An intuitive way to perceive the squeezing phenomena is to plot the
Wigner function
\begin{equation}\label{wigner}
\rho_W(x',p')=(2\pi)^{-1}\int_{-\infty}^{\infty} \langle
x'-\frac{1}{2}y|\rho|x'+\frac{1}{2}y\rangle e^{ip'y}\,dy.
\end{equation}
The Wigner function is one of the quantum-mechanical analogues to
the phase space probability distribution. Even though it is not a
genuine 2-dimensional probability distribution, it has the property
of giving the one-dimensional marginal distributions in arbitrary
directions in the phase space $(x',p')$ and can therefore be used
directly to illustrate the uncertainties of observables. The
groundstate of the SQUID has a circular symmetry which is distorted
to an ellipse by squeezing, see Fig. \ref{wig1}. For ideal squeezed
states the principal axes of elliptical contour lines are inversely
proportional reflecting the minimum uncertainty property. The
primary effect of the dissipation is to broaden the shorter axis
resulting in an excess uncertainty.

\section{spectrum of parametric resonance and squeezing}

Based on Eq. (\ref{voltage2}) the spectral properties of $V(X,t)$
are directly related to spectral properties of $\phi$ and the free
TL spectrum. Here we evaluate $\phi$-noise and, at the end of the
section, discuss parameters relevant to the TL output noise.
\subsection{Time-averaged spectrum }
The spectrum contains information about radiation emitted from the
studied system. In Ref. \onlinecite{schoelkopf} an excitation of a
two-level system with an energy splitting $\Omega=\Delta E$ is
considered as a spectrum analyzer. The probability to detect the
analyzer system in either of its states is proportional to
\begin{equation}\label{double}
\int_0^t\int_0^t<A(t_1)A(t_2)>e^{-i\omega(t_2-t_1)}dt_1dt_2
\end{equation}
at $\omega=\pm\Omega$, where $A$ is an operator of the studied noise
source which couples to the analyzer two-level system. Supposing
that the system is in a steady state and the correlation time is
short, the expression ($\ref{double}$) reduces to $tS_A(\omega)$,
where
\begin{equation}\label{spec1}
S_A(\omega)=\int_{-\infty}^{\infty}
<A(\tau)A(0)>e^{i\omega\tau}d\tau.
\end{equation}
 The expression (\ref{spec1}) is
positive definite and determines the transition rates of the
analyzer system. In the strong damping regime the SQUID radiation is
periodic and the steady-state expression (\ref{spec1}) is not
appropriate as such. Rather, the detector rates are proportional to
the time-averaged noise power
\begin{equation}\label{spec2}
S_A(\omega)=\int_{-\infty}^{\infty}\frac{1}{T}\int_{0}^T
<A(\tau+t')A(t')>e^{i\omega\tau}dt'd\tau,
\end{equation}
where $T$ is the period of the motion. The expression (\ref{spec2})
generalizes (\ref{spec1}) in the sense that so defined spectrum is
positive definite and gives excitation probabilities of the analyzer
system in the long-time limit $t\gg T=2\pi/\omega_0$. Since the
SQUID couples to the TL with the operator
$\phi'=(a+a^{\dagger})/\sqrt{2}$, the $\phi'$-spectrum is relevant
in the TL radiation.

\begin{figure}[h]
\centering
\begin{picture}(100,260)
\put(-30,250){(a)}
\put(-20,130){\includegraphics[width=0.60\columnwidth]{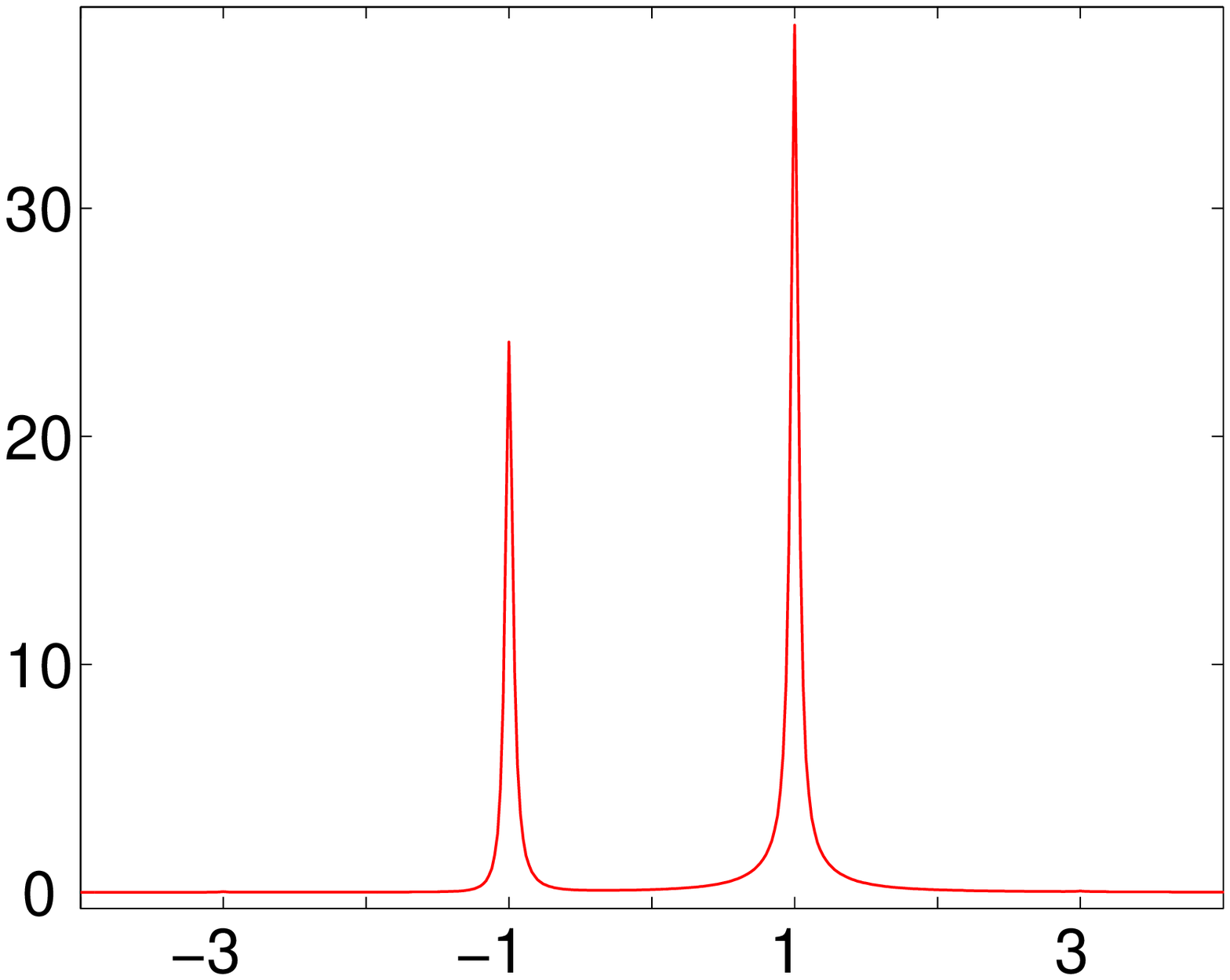}}
\put(95,145){$\omega\,[\omega_0]$} \put(0,235)
{$S_{\phi'}(\omega)\,[\omega_0^{-1}]$} \put(-30,120){(b)}
\put(-20,0){\includegraphics[width=0.60\columnwidth]{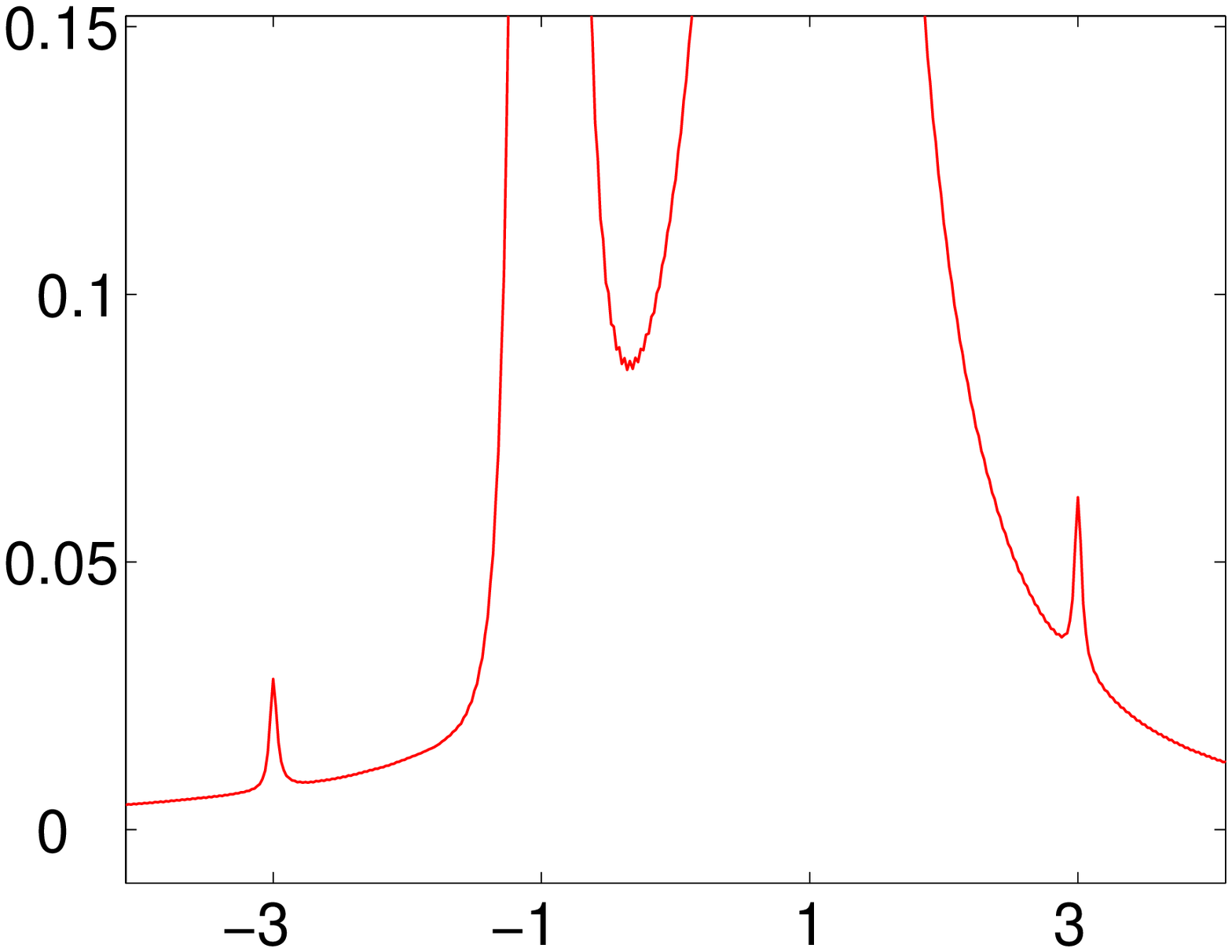}}
\put(95,15){$\omega\,[\omega_0]$} \put(0,100)
{$S_{\phi'}(\omega)\,[\omega_0^{-1}]$}
\end{picture}
\caption{(Color online) (a)Time-averaged noise spectrum
$S_{\phi'}(\omega)$
 of the periodic squeezed state $\kappa=0.15\omega_0=1.25B$. The Spectrum has resonances at frequencies
 $\pm\omega_0$. (b) The fine structure of spectrum (a) reveals the weak resonance peaks at
 $\pm3\omega_0$. }
\label{spect1}
\end{figure}

From the time-averaged observables it is difficult to detect the
squeezing directly. This can be understood by considering the Wigner
function. The fast rotation of the elliptic figure in Fig.
\ref{wig1} averages out to a circular shape where no squeezing is
present. The $\phi$-spectrum (\ref{spec2}) of the SQUID in
parametric resonance resembles to a spectrum of a dissipative
harmonic oscillator at finite temperature, see Fig. \ref{spect1}.
The spectrum displays peaks at frequencies $-\omega_0$ and
$\omega_0$ corresponding to emission and absorption processes. A
study of a fine structure also reveals weak resonance peaks at
frequencies $-3\omega_0$ and $3\omega_0$. These are the higher
harmonics produced by the parametric driving process. In principle
there is also resonance at higher odd multiples of $\omega$ but they
are extremely weak compared to $-\omega_0$ and $\omega_0$ peaks. The
distance between the peaks is $2\omega$ which characterizes the
periodicity of the rotating squeezing.

\subsection{Spectrum of rotating operators}
\begin{figure}[h]
\centering
\begin{picture}(100,150)
\put(-30,0){\includegraphics[width=0.70\columnwidth]{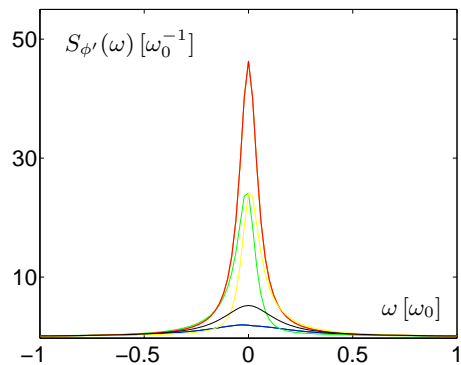}}
\put(110,20){$\omega\,[\omega_0]$} \put(-10,120)
{$S_{\phi'}(\omega)\,[\omega_0^{-1}]$}
\end{picture}
\caption{(Color online) Noise spectrum $S_{\phi'}(\omega)$ of the
squeezed state $\kappa=0.15\omega_0=1.5B$ in the rotating operators.
The resonance peaks are now moved to the origin. The initial time is
chosen so that the minimum noise curve (blue) corresponds to
$\theta=0$. The green, red and yellow curves correspond to the
orientations $\theta=\pi/4,\pi/2,3\pi/4$ respectively. The black
curve represents the groundstate noise. The minimum noise curve
remain below the groundstate noise, which is an evidence of
squeezing.} \label{rotspect1}
\end{figure}

The detailed analysis of squeezing can be performed in a
phase-sensitive measurement. For that purpose we define the rotating
frame operators as $b(t)=\mathrm{exp}(i\omega_0t+\theta)a(t)$ and
$b^{\dagger}(t)= \mathrm{exp}(-i\omega_0t-\theta)a^{\dagger}(t)$.
The rotating SQUID observables are defined analogously as
$\phi'=(b(t)+b^{\dagger}(t))/\sqrt{2}$ and
$Q'=i(b^{\dagger}(t)-b(t))/\sqrt{2}$. The rotating phase factors in
 $b(t)$ and $b^{\dagger}(t)$  compensate the
natural rotation of the harmonic creation and annihilation
operators. In the new operators the orientation of the squeezing is
almost static and determined by the angle $\theta$. The Wigner
functions in Fig. \ref{wig1} appear frozen in time. Only a tiny
time-dependent deformations resulting from the driving remains. The
noise varies periodically as a function of $\theta$ as the
orientation of the squeezing is rotated, see Fig. \ref{rotspect1}.
For a finite range of $\theta$ the SQUID noise goes below its
groundstate value which is a direct evidence of reduction of the
quantum fluctuations. For the parameters of Fig. \ref{rotspect1},
the groundstate noise at zero frequency is approximately 2.5 times
higher than the minimum noise corresponding to $\theta=0$. The noise
curves of squeezed states are slightly distorted respect to the
origin, which indicates that there is a tiny asymmetry respect to
the half axes of the ellipse in Fig. \ref{wig1}. A parametric
driving in the presence of a dissipation does not lead to a
completely symmetric squeezing.

A phase-sensitive measurements can be realized, for example, with a
help of a parametric amplification process.\cite{yurke, yurke1} A
harmonic modulation of the measurement signal with the frequency
$\omega_0$ can be used to measure the rotating
spectrum.\cite{braginsky}

\subsection{Transmission line output}
The TL observables are sum of a free line observables and a retarded
SQUID radiation contribution, see Eq. ($\ref{voltage2}$). The
voltage noise of the TL is of the form
$S_{V}(\omega)=S_{V}^{0}+g^2S_{\phi'}(\omega)$, where $S_{V}^{0}$ is
the vacuum noise of the TL. The coupling constant $g$ depends on the
actual material parameters and can be deduced from Eq.
(\ref{voltage2}):
\begin{equation}\label{coupl}
g=\frac{M\omega_0}{L_S}\sqrt{\frac{\hbar Z_0}{2}},
\end{equation}
where $Z_0=\sqrt{L_S/C}$. For the phenomenon to be measurable, the
vacuum noise of the TL should not overwhelm the SQUID noise. From
Eq. $\ref{coupl}$ we can estimate the order of magnitude as
$g^2S_{\phi'}(\omega_0)\approx (M/L_S)^2\hbar\omega_0 Z_0$. At low
temperatures the free TL vacuum noise is
$S_{V}^{0}=\hbar\omega_0Z_{\mathrm{TL}}$, where
$Z_{\mathrm{TL}}=\sqrt{l/c}$ is the characteristic impedance of the
TL. Then the relative effect is of order
$g^2S_{\phi'}(\omega_0)/S_{V}^{0}(\omega_0)=(M/L_S)^2Z_0/Z_{\mathrm{TL}}$.
For realistic values $M/L_S=0.2$, $Z_0=500\Omega$ and
$Z_{\mathrm{TL}}=50\Omega$ the fraction becomes
$g^2S_{\phi'}(\omega_0)/S_{V}^{0}(\omega_0)=0.4$, which suggests
that the SQUID noise is significant contribution to the TL output
radiation in favorable conditions.

\section{conclusions}
The parametric harmonic driving creates rotating squeezed quantum
states in the SQUID ring. If the environmental damping of the SQUID
is negligible the squeezing of uncertainties is exponential for
short times. In the presence of a strong damping the magnitude of
the squeezing is stationary. The minimum uncertainty in the flux and
charge go below the groundstate value periodically. The phenomenon
enables a quantum noise engineering which plays an increasingly
important role in the quantum measurement theory as well as in the
design of practical quantum devices. In principle, squeezed harmonic
systems could be used as an ultra sensitive measurement devices of
external perturbations coupled to it.\cite{ruskov,braginsky}

The experimental creation of squeezed quantum states in a SQUID by a
parametric driving is feasible with the current experimental
methods. We calculated the relevant noise properties of the periodic
squeezed state. By introducing a coupling to the transmission line,
we analyzed the fully quantum mechanical emission spectrum of the
SQUID and discussed briefly the conditions of an experimental
verification of the phenomenon.

\end{document}